# Theoretical and numerical analysis for angular acceleration being determinant of brain strain in mTBI

Yuzhe Liu[1,*,#], Xianghao Zhan[1,*], August G. Domel[1], Michael Fanton[2], Zhou Zhou[1], Samuel J. Raymond[1], Hossein Vahid Alizadeh[1], Nicholas J. Cecchi[1], Michael Zeineh[4], Gerald Grant[5]

[1] Department of Bioengineering, Stanford University, Stanford, CA, 94305, USA.

[2] Department of Mechanical Engineering, Stanford University, Stanford, CA, 94305, USA.

[3] Department of Neurosurgery, Stanford University, Stanford, CA, 94305, USA.

[4] Department of Radiology, Stanford University, Stanford, CA, 94305, USA.

[5] Department of Neurology, Stanford University, Stanford, CA, 94305, USA.

* These authors contribute equally to this study.

[#] Corresponding author (e-mail: yuzheliu@stanford.edu)

## Abstract

Mild traumatic brain injury (mTBI, also known as concussion) caused by head impact is a crucial global public health problem, but the physics of mTBI is still unclear. During the impact, the rapid movement of the head injures the brain, so researchers have been endeavoring to investigate the relationship between head kinematic parameters (e.g., linear acceleration, angular velocity, angular acceleration) and brain strain, which is associated with the injury of the brain tissue. Although previous studies have shown that linear acceleration had limited contribution to brain strain, whether angular velocity or angular acceleration causes brain strain is still unclear because of their interdependency (acceleration being the velocity's time-derivative). By reframing the problem through the lens of inertial forces, we propose to use the skull frame of reference instead of the ground frame of reference to describe brain deformation during head impact. Based on the rigid-body rotation of the brain, we present a theoretical framework of mechanical analysis about how the inertial forces cause brain strain. In this way, we theoretically show that angular acceleration determines brain strain, and we validate this by numerical simulations using a finite element head model. We also provide an explanation of why previous studies based on peak values found the opposite: that angular velocity determined brain strain in certain situations. Furthermore, we use the same framework to show that linear acceleration causes brain strain in a different mechanism from angular acceleration. However, because of the brain's different resistances to compressing and shearing, the brain strain caused by linear acceleration is small compared with angular acceleration.

Mild traumatic brain injury (mTBI, **Table 1** gives abbreviation and symbols) is a crucial global public health problem because it significantly impairs patients' quality of life [1,2]. Because of the relatively thick skull of humans, the stress wave caused by a head impact will not pass through the skull [3]. Instead, the head impact will cause a rapid movement of the head, which in turn deforms the brain because of its inertia. Therefore, head kinematics are the key to describing the severity of the loading on the brain. The traumatic brain injury community has endeavored to learn how head kinematics predict brain strain, a mechanical parameter describing deformation and closely relevant to brain injury [4,5]. In 1943, Holbourn hypothesized that linear acceleration would not deform the brain significantly and was unlikely to cause brain injury because of the incompressibility of the brain [6]. After that, researchers still debated whether translation or rotation of the head accounted for brain injury [7-9] until the recent development of finite element (FE) head models [3,10-12] which could calculate brain strain according to head kinematics. The simulation using a FE head model helped confirm the Houlbourn hypothesis that brain strain was most dependent on head rotation instead of translation [3], but the mechanism behind this has not been formally discussed.

Angular acceleration and angular velocity of the head have been studied heavily over the last two decades to better understand how the brain is deformed [13-17]. Based on these theories, reduced-order models were developed to predict brain strain. The effects of linear and angular acceleration have been combined to predict brain strain [18-22]. In contrast, angular velocity was used [12,23-25] and found to have better predictability for the peak brain strain [12,26]. According to the studies comparing reduced-order models [12,24,27], the angular velocity-based models have achieved a higher predictability than angular acceleration, indicating that the angular velocity decides brain strain. However, the better predictability of angular velocity contradicted both experimental and numerical results [28], where volunteers experienced angular velocity higher than the injury threshold [13], but had no brain injury and low brain strains. This finding [28] could be explained by the classification of the impacts according to the impulse duration [16]: the peak brain strain depends on angular velocity peak in short-duration impacts, on angular acceleration peak in long-duration impacts and on both in moderate-duration impacts. However, the mechanism behind this classification [16] is still unclear. Furthermore, it should be noted that only peak values of kinematic parameters and brain strain were investigated [11-13,23-25,27], and the angular velocity with different profiles but the same peak were found to yield different brain strains [29-31]. Recently, combining the effects of both angular velocity and acceleration, researchers developed mass-spring-damper models [16,17,32,33] and machine learning models [26,34,35] to predict brain strain and have achieved promising accuracy. However, the mechanisms underlying the relationship between brain strain and head kinematics are still unknown.

The difficulty in separating the effects of the head angular velocity and acceleration is the interdependency between them, that is, angular acceleration is the derivative of angular velocity. The movement of the head in the ground frame of reference (FoR) will be determined when each of them is given (**Fig.1A**, the FoR is fixed to the ground). Therefore, the brain strain caused by angular acceleration and angular velocity can not be investigated independently. To address this, we studied brain deformation in the skull FoR (**Fig.1B**, the FoR moves along with the skull). In the skull FoR, the skull is fixed, and the brain tissue is deformed by the inertial forces caused by each kinematic parameter as shown in **Fig.1B**. In this manner, the effect of each kinematic parameter can be investigated independently.

To explain how each inertial force causes brain strain, the rigid-body movement of the brain [36] found in low-severity cadaver head impacts [37] was used as a bridge. The rigid-body movement of the brain means that most of the brain tissue is displaced similarly as a rigid body. However, because of the constraint of the skull and the spatial distribution of the inertial force, brain strain will still be caused. In this study, we assumed that brain strain is linearly correlated with the magnitude of the rigid-body rotation of the brain (RRB) and validated this assumption by FE simulations.

In this letter, we show that only the inertial force by angular acceleration causes non-zero torque on the brain (**Fig.1C**), therefore only angular acceleration leads to the RRB and the brain strain. Then, to validate this finding, we performed FE simulations using the KTH head model [11] and 118 on-field football head impacts (**Fig.2**) collected by instrumented mouthguards [38-40]. Considering the difference between our findings and previous studies [12,16,24,27], we provide a further explanation based on the brain strain by constant angular acceleration. Moreover, we discuss the rigid-body translation of the brain (RTB) caused by linear acceleration, which provides further explanation to the Holbourn hypothesis [6].

The skull FoR is non-inertial, and the inertial force (Eq.1) caused by the movement of the FoR is [41],

$$\boldsymbol{F}(\boldsymbol{r}) = \rho(\boldsymbol{r})\boldsymbol{a} - \rho(\boldsymbol{r})\boldsymbol{\omega} \times (\boldsymbol{\omega} \times \boldsymbol{r}) - \boldsymbol{\beta} \times \boldsymbol{r} - 2\rho(\boldsymbol{r})\boldsymbol{\omega} \times \boldsymbol{v}_r(\boldsymbol{r}) \qquad (\text{Eq.1})$$

Where $\boldsymbol{F}(\boldsymbol{r})$ is the inertial force per unit mass at the location $\boldsymbol{r}$; $\boldsymbol{a}$, $\boldsymbol{\omega}$ and $\boldsymbol{\beta}$ are the linear acceleration at coordinate origin, the angular velocity, and the angular acceleration of the head, respectively; $\boldsymbol{v}_r(\boldsymbol{r})$ is the relative velocity of location $\boldsymbol{r}$ at the skull FoR; $\boldsymbol{r}$ is the position vector, and the coordinate origin of $\boldsymbol{r}$ is set at the center of gravity (CoG) of the brain (**Fig.1A**), therefore,

$$\iiint \boldsymbol{r}\rho(\boldsymbol{r})\mathrm{d}v = \boldsymbol{0} \qquad (\text{Eq.2})$$

Because the relative movement of the brain tissue in the skull was small [36,37], the Coriolis force item ($2\rho(\boldsymbol{r})\boldsymbol{\omega} \times \boldsymbol{v}_r(\boldsymbol{r})$) can be neglected, and this is validated by FE simulation later. Therefore, Eq.1 is simplified as Eq.3. It should be noted that the items in Eq.3 with $\boldsymbol{a}$, $\boldsymbol{\omega}$ and $\boldsymbol{\beta}$ are separated, so the effect of each kinematic parameter can be studied independently. We defined the inertial forces by each kinematic parameter as,

$$\begin{aligned}\boldsymbol{F}(\boldsymbol{r}) &= \rho(\boldsymbol{r})\boldsymbol{a} - \rho(\boldsymbol{r})\boldsymbol{\omega} \times (\boldsymbol{\omega} \times \boldsymbol{r}) - \boldsymbol{\beta} \times \boldsymbol{r} \\ &= \boldsymbol{F}_{\text{LinAcc}}(\boldsymbol{r}) + \boldsymbol{F}_{\text{AngVel}}(\boldsymbol{r}) + \boldsymbol{F}_{\text{AngAcc}}(\boldsymbol{r})\end{aligned} \qquad (\text{Eq.3})$$

Where $\boldsymbol{F}_{\text{LinAcc}}$, $\boldsymbol{F}_{\text{AngVel}}$ and $\boldsymbol{F}_{\text{AngAcc}}$ are the inertial forces by linear acceleration, angular velocity, angular acceleration, respectively (**Fig.1B**), and are defined as,

$$\boldsymbol{F}_{\text{LinAcc}}(\boldsymbol{r}) = \rho(\boldsymbol{r})\boldsymbol{a} \qquad (\text{Eq.4})$$

$$\boldsymbol{F}_{\text{AngVel}}(\boldsymbol{r}) = -\rho(\boldsymbol{r})\boldsymbol{\omega} \times (\boldsymbol{\omega} \times \boldsymbol{r}) \qquad (\text{Eq.5})$$

$$\boldsymbol{F}_{\text{AngAcc}}(\boldsymbol{r}) = -\rho(\boldsymbol{r})\boldsymbol{\beta} \times \boldsymbol{r} \qquad (\text{Eq.6})$$

Integrating over the brain, the inertial torque on the brain caused by the head movement is calculated at the CoG of the brain,

$$\boldsymbol{T} = \iiint \boldsymbol{r} \times \boldsymbol{F}(\boldsymbol{r})\mathrm{d}v \qquad (\text{Eq.7})$$

Replacing $F(r)$ by Eq.3, the inertial torque can be written as the sum of the inertial torques caused by each kinematic parameter,

$$T = T_{\text{LinAcc}} + T_{\text{AngAcc}} + T_{\text{AngAcc}} \tag{Eq.8}$$

Here, the inertial torque by linear acceleration is,

$$T_{\text{LinAcc}} = \iiint r \times F_{\text{LinAcc}}(r) \text{dv} \tag{Eq.9}$$

Replacing Eq.4 into Eq.9 and considering $a$ is constant over the brain,

$$T_{\text{LinAcc}} = \iiint \rho(r) r \times a \text{dv} = \iiint \rho(r) r \text{dv} \times a = 0 \tag{Eq.10}$$

Then, the inertial torque by angular velocity is,

$$T_{\text{AngVel}} = \iiint r \times F_{\text{AngVel}}(r) \text{dv} \tag{Eq.11}$$

Replacing Eq.5 into Eq.11,

$$T_{\text{AngVel}} = \iiint \rho(r)(\boldsymbol{\omega} \cdot r) \boldsymbol{\omega} \times r \text{dv} \tag{Eq.12}$$

Further, the inertial torque by angular acceleration is

$$T_{\text{AngAcc}} = \iiint r \times F_{\text{AngAcc}}(r) \text{dv} \tag{Eq.13}$$

Replacing Eq.6 into Eq.13,

$$T_{\text{AngAcc}} = \iiint \rho(r)(r^2 \boldsymbol{\beta} - (r \cdot \boldsymbol{\beta}) r) \text{dv} \tag{Eq.14}$$

We further assume that the shape of the human brain is close to a sphere (radius $r_0$) and has a homogeneous density $\rho_0$. Because of the symmetry in Eq.12, the inertial torque by angular velocity is,

$$T_{\text{AngVel}}^{\text{sphere}} = 0 \tag{Eq.15}$$

Applying the sphere assumption to Eq.14, we get the inertial torque by angular velocity for a spherical brain as (see deduction of Eqs. 15 and 16 in Supplementary Information S1),

$$T_{\text{AngAcc}}^{\text{sphere}} = \frac{8}{15} \rho_0 \pi r_0^5 \boldsymbol{\beta} \tag{Eq.16}$$

Therefore, for the actual brain,

$$T_{\text{AngVel}} \ll T_{\text{AngAcc}} \tag{Eq.17}$$

The accuracy of Eq.17 depends on the closeness of the human brain to the sphere. To validate Eq.17, the inertial toques in 118 on-field football impacts (**Fig.2**) were calculated for the brain shape of the KTH head model. As shown in **Fig.1C**, the results agree with Eqs.10 and 17.

Therefore, the inertial torque is only decided by the angular acceleration. Eq.8 can be rewritten as,

$$T = T_{\text{AngAcc}} \quad \text{(Eq.18)}$$

At the skull FoR, the RRB is decided by two torques: the inertial torque ($T$) and the constraint torque by the skull ($T_{\text{skull}}$), which is caused by the constraint between the rotating brain and the fixed skull. Assuming the brain is static at the skull FoR before the impact, based on the conservation of angular momentum of the brain,

$$\int (T + T_{\text{skull}}) \mathrm{d}t = L \quad \text{(Eq.19)}$$

Where $L$ is the angular momentum of the brain at the skull FoR. Because $T_{\text{skull}}$ is caused by the constraint, its value is decided by the brain rotation. Therefore, $T_{\text{skull}}$ is a function of $L$, and Eq.19 can be transformed as,

$$\int T \mathrm{d}t = L - \int T_{\text{skull}}(L) \mathrm{d}t \quad \text{(Eq.20)}$$

It should be noted that $T_{\text{skull}}(L)$ is a function determined by the head. Therefore, according to Eq.20, $L$ is only decided by $T$. As shown in Eq.18, the angular acceleration determines inertial torque on the brain, and thus determines the angular momentum of the brain in head impact. Then, we define the equivalent angular velocity of the brain ($\dot{\phi}$) to describe the RRB,

$$L = I_C \dot{\phi} \quad \text{(Eq.21)}$$

Where $I_C$ is the moment of inertia about the instantaneous axis of the rotation which passes the CoG of the brain, which can be calculated as shown in Supplementary Information S2. Then, the equivalent angle of the brain ($\phi$) is defined as the magnitude of the integration of $\dot{\phi}$ using quaternion [42]. Since $\phi$ is calculated from the angular momentum $L$, it represents the magnitude of RRB.

As mentioned above, in low-severity head impacts, the movement of brain tissue can be described as a rigid body [36]. Therefore, $\phi$ (indicating the magnitude of RRB) and brain strain both describe the severity of brain deformation and should have similar variation. Therefore, we assume $\phi$ is linearly correlated to the 95[th] percentile maximum principal strain (95% MPS). To validate this, we extracted $\phi$ and 95% MPS from FE simulations with 118 on-field football head impacts, and calculated Pearson correlation coefficients. As shown in **Fig.4A**, most of the correlation coefficients were higher than 0.95 and the lowest was 0.88. The head impacts with the highest 95% MPS peak (**Fig.4B**) and lowest correlation (**Fig.4C**) are also shown as examples. This result shows that $\phi$ is linearly correlated with 95% MPS. As shown in **Fig.1C**, $\phi$ is calculated by the angular momentum (Eqs. 20-21), which is only decided by angular acceleration (Eqs. 18 and 20). Therefore, angular acceleration determines 95% MPS.

To validate this result, we simulated 118 on-field football head impacts at the skull FoR. As shown in **Fig.5A**, the simulations with all inertial forces items (Eq.1) at the skull FoR give the same results as the simulations with the traditional loading (moving head at the ground FoR). The impacts with the highest peak (**Fig.5B**), the largest difference in the peaks (0.009, **Fig.5C**), and the lowest Pearson correlation coefficient (0.80, **Fig.5D**) were given as examples. Then, we performed simulations of inertial force by individual kinematic parameter (Eqs.4-6) and compared the results with that of inertial force by all kinematic parameter (Eq.3) in **Figs.6A and C**. The 95% MPS traces and MPS distribution in the case with the highest 95% MPS peak were plotted in **Figs.6B and D-G**. It is clear that 95% MPS caused by $F_{\text{AngAcc}}$ is almost the same as actual 95% MPS ($F_{\text{LinAcc}}(r) + F_{\text{AngVel}}(r) + F_{\text{AngAcc}}(r)$). This result validates that angular acceleration determines 95% MPS, and the effect of linear acceleration (Eq.4), angular velocity (Eq.5), and Coriolis force are negligible. This result also explains the experimental and numerical results that no injury was observed with high angular velocity and low angular acceleration [28]. Considering angular acceleration is decided by angular velocity profiles, this finding also explains that the angular velocities with the same peak but different profiles yielded different 95% MPS peak [31].

The above finding raises the following questions: (1) Why do the angular velocity peak and the reduced-order models based on the angular velocity peaks have better predictability of brain strain [11-13,24,25,27]? The data we used in the validation also shows that 95% MPS correlates better with angular velocity (**Fig.2B**, $R^2 = 0.70$) than angular acceleration (**Fig.2A**, $R^2 = 0.61$); (2) Why does 95% MPS depend on different kinematic parameters according to the duration of the impact [16]?

To explore these questions, we performed simulations in which the angular accelerations are square waves with impulse durations from $1\text{ ms}$ to $15\text{ ms}$. In these simulations, constant angular accelerations of $8, 16, 24\text{ krad/s}^2$ in X, Y, Z directions were applied independently. Then, the angular velocity was calculated according to the angular acceleration (assuming static at the beginning) and the linear acceleration was set to zero. In each direction, as shown in **Fig.7**, when the impulse duration is short, the 95% MPS peak increases linearly with the impulse duration (data points were fitted to the red dash-dot linear lines, $R^2$>0.997 for each red line). Since angular velocity is the integration of angular acceleration, the angular velocity peak also increases linearly with the impulse duration. As a result, the angular velocity peak correlates well with the 95% MPS peak. However, this correlation is owing to that both angular velocity and 95% MPS have a linear correlation with the impulse duration and does not necessarily indicate that brain strain is caused by the angular velocity. Then, as the impulse duration gets longer, the 95% MPS peak increases nonlinearly with the duration, and is decided by the combination of angular acceleration and angular velocity. In long-duration impacts, the 95% MPS peak keeps at a constant level decided by the angular acceleration. The dependency of 95% MPS peak on the impulse duration agrees with the findings by Gabler et al. [16], although different types of idealized loadings (sine function [16]) and different FE head models (GHBMC [16]) were used. It should be noted that the impulse duration was the length of square waves, which is different from the duration of real head impacts. Therefore, the value of the impulse duration in **Fig.7** should not be directly applied to the analysis of on-field data. It was suggested that most types of head impacts, including sled, crash, and sports accidents, fell into the region of moderate-duration impulse [16]. Further studies are needed to define the impulse duration for real head impacts.

The different regions in **Fig.7** and [16] also showed that the head impact should be evaluated by entire traces rather than peak values alone. It was also suggested that different 95% MPS peaks were observed with the same angular velocity peaks but different traces of angular acceleration [29-31]. Moreover, although the peak values of 95% MPS were always used to predict brain injury [16,17,32,35], the trace of 95% MPS may also play an important role on the tissue pathology: the damage accumulates when the brain sustains large deformation during the impact. Although no direct neurological data supports this to the best of our knowledge, the accumulation effect of mTBI in repeated head impacts has been suggested [43-47]. Therefore, although angular velocity peak correlates well with the 95% MPS peak in several types of head impacts, a comprehensive evaluation of head impact should be based on the entire trace of angular acceleration, which requires accurate measurement of it. Recently, an algorithm to compute accurate angular acceleration without derivative was developed [48], which shed light on monitoring mTBI based on the accurate measurement of angular acceleration in the future.

The idea of simplifying brain deformation as RRB was adopted in reduced-order models predicting brain strain by kinematics ([15-17,32,33]), and $T_{\text{skull}}$ was simplified as spring and damper in these models. Compared with previous reduced-order models, the models based on the assumption of RRB manifested promising accuracy in predicting brain strain across different types of head impacts [49]. The analyses of RRB and bran strain in this letter explains that the high accuracies of these reduced-order models are results of that the brain displaces as a rigid-body. Additionally, it should be noted that the brain strain contributed by the rigid-body movement and local deformation were comparable in the milder head rotation (angular acceleration: $250 \sim 300 \text{ rad/s}^2$, [50,51]). However, the head kinematics that led to mTBI [5,52-54] are more severe than those observed in volunteer experiments [50,51], and are close to low-severity head impacts [36] where the rigid-body movements were observed (**Table.2**). Therefore, we adopted the RRB assumption that the brain is mainly displaced as a rigid body [36].

Recently, the same head kinematics yielded different traces of 95% MPS for different brains [55], which indicated the necessity of customized brain injury criterion to evaluate injury risk. In Eq.19, the interaction between the skull and the brain is represented by $T_{\text{skull}}$, which should be decided by the geometry of the brain and the properties of cerebrospinal fluid [56]. The influence of the brain size is also represented in $I_C$ in Eq.21 and the integration region for $T_{\text{AngAcc}}$ in Eq.14. $T_{\text{skull}}$, $I_C$ and the integration varies from person to person. Further investigation of the influence of individual differences is warranted.

In Eq.10, $T_{\text{AngAcc}}$ is zero, therefore the RRB is irrelevant with the linear acceleration. However, it should be noted that $F_{\text{LinAcc}}$ caused low but non-zero 95% MPS (**Fig.6**). This indicates that the linear acceleration deforms the brain via another mechanism different from the RRB. In [36], besides the RRB, the RTB was also observed. While RRB deforms brain by shearing the tissue, RTB can also deform the brain by stretching and compressing brain tissue. Assuming the brain is static at the skull FoR before the impact, based on the conservation of momentum of the brain,

$$\int (R_{\text{AngAcc}} + R_{\text{AngVel}} + R_{\text{LinAcc}} + R_{\text{skull}}) \cdot \text{dt} = P \tag{Eq.23}$$

Where $P$ is the momentum of the brain and $R_{\text{skull}}$ is the resultant force on the brain by the skull. Similar to the analysis for $T_{\text{skull}}$, $R_{\text{skull}}$ is a function of the history of $P$. $R_{\text{AngAcc}}$, $R_{\text{AngVel}}$, and $R_{\text{LinAcc}}$ are the resulting inertial forces by each kinematic parameter,

$$R_{\text{AngAcc}} = \iiint F_{\text{AngAcc}}(r) \text{dv} \tag{Eq.24}$$

$$R_{\text{AngVel}} = \iiint F_{\text{AngVel}}(r) \text{dv} \tag{Eq.25}$$

$$R_{\text{AngVel}} = \iiint F_{\text{LinAcc}}(r) \text{dv} \tag{Eq.26}$$

Replacing Eqs.3-6 into Eqs 24-26,

$$\boldsymbol{R}_{\text{AngAcc}} = \boldsymbol{0} \tag{Eq.27}$$

$$\boldsymbol{R}_{\text{AngVel}} = \boldsymbol{0} \tag{Eq.28}$$

$$\boldsymbol{R}_{\text{AngVel}} = m\boldsymbol{a} \tag{Eq.29}$$

Where m is the brain mass. Replacing Eqs.27-29 into Eq.23,

$$m\int \boldsymbol{a} \cdot \mathrm{dt} = \boldsymbol{P} - \int \boldsymbol{R}_{\text{skull}}(\boldsymbol{R})\mathrm{dt} \tag{Eq.30}$$

Eq.30 shows that the linear acceleration will determine the RTB, which will further cause brain strain by pulling and pressing tissue. However, because the bulk modulus of the brain is significantly higher than the shearing modulus [6,11,57], the MPS caused by RTB is negligible compared with that by RRB. This explains that the low but non-zero MPS by $F_{\text{LinAcc}}$ observed in **Fig.6**. Furthermore, it should be noted that the locations of the 95% MPS peak caused by RTB are unlikely to be the same as the locations of that caused by RRB, therefore the actual 95% MPS peak is decided by the higher of the two, that is, RRB. This agrees with the Holbourn hypothesis [6] that linear acceleration has negligible effect on brain strain and the simulation study about the effect of linear acceleration [3]. However, the RTB will slightly influence the distribution of MPS for the whole brain (**Fig.6F**).

This letter has several limitations: Although the FE simulations confirmed that the 95% MPS and equivalent angle of the brain $\phi$ are linearly correlated (**Fig.4**), $\phi$ and 95% MPS may not correspond to each other at every time point because the mechanical properties of the brain tissue are highly history-dependent [57,58]. Another limitation is that the relationship between the brain strain and the inertial force by each kinematic parameter (Eq.3) is not linear. Therefore, the actual brain strain is not the same as the sum of the brain strain caused by each individual kinematic parameter, and we could not compute the ratio of the contribution to brain strain by each kinematic parameter. Furthermore, $T_{\text{AngVel}}$ is zero (Eq.15) when the brain is assumed as a sphere. This assumption is based on the geometry of the human brain, and should be reconsidered with the animal brain in preclinical studies [59]. We also assumed that the Coriolis force is neglected since the relative velocity of brain to skull is small and validated this assumption by the FE simulations with on-field data (**Fig.6**). However, the effect of the Coriolis force should be inspected in more severe head impacts since the brain tissue may move faster.

In summary, we propose an alternative perspective to describe brain deformation in mTBI. Contrary to previous studies in which the brain was deformed by the skull movement in the ground FoR (inertial), we describe this process as the brain being deformed by the inertial force in the skull FoR (non-inertial). This perspective allows us to separate the effects of each individual kinematic parameter. Then, using the rigid-body rotation of the brain (RRB) as a bridge, we find that 95% MPS is only determined by angular acceleration instead of angular velocity, and validate this by FE simulations in the skull FoR. Considering that previous studies found that the angular velocity decided brain strain [12,16,17,24,27], we provide an explanation with FE simulations, and show that this is owing to the linear relationship between brain strain and the impulse duration. Furthermore, we find two independent mechanisms of brain deformation: the RRB decided by angular acceleration and rigid-body translation of the brain (RTB) decided by linear acceleration. Because of the brain tissue properties, the brain strain caused by the RTB is negligible compared with that by the RRB, and this provides further explanation to the Holbourn hypothesis [6].

This letter shows the importance of the angular acceleration during a head impact to resulting mTBI. Therefore, when designing and evaluating wearable devices that measure head impact kinematics, extra attention should be given to the accuracy of angular acceleration measurement. Furthermore, the traces of angular acceleration, instead of other kinematic parameters, should be used as input for future reduced-order models to estimate brain strain and predict concussion risk. This study suggests the potential to extract a pattern of brain deformation based on RRB and provides a theoretical framework of mechanical analysis about how brain strain is caused by head impact: the loading is represented by the inertial torque, the interaction between brain and skull is represented by the $T_{\text{skull}}$, and the resulting RRB decides the brain strain. These three parts can be future research directions to further understand how the shape, size, and material property of the brain influence its strain responses.

This research was supported by the Pac-12 Conference's Student-Athlete Health and Well-Being Initiative, the National Institutes of Health (R24NS098518), Taube Stanford Children's Concussion Initiative. We also want to thank Dr. Svein Kleiven for sharing the finite element head model, and Dr. Yue Gao for the insightful discussion.

*Yuzhe Liu and Xianghao Zhan contributed equally to this work.

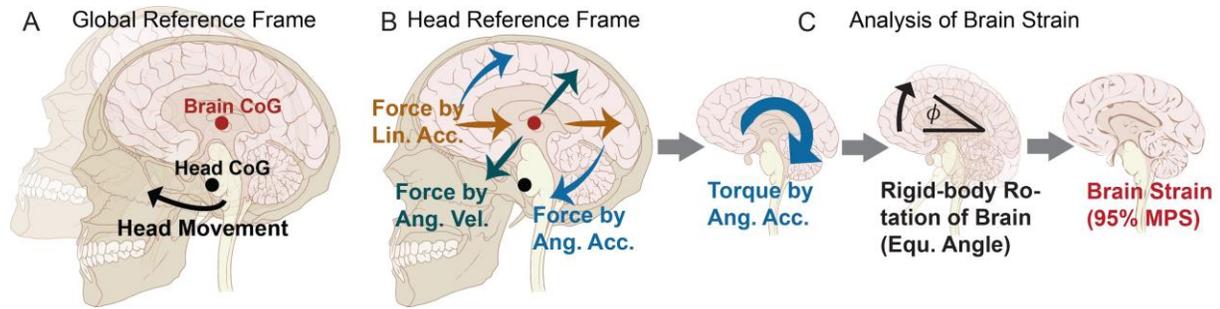

Figure 1. The diagrams of how head impact causes the brain strain at (A) the ground frame of reference (FoR) and (B) the skull FoR. The center of gravity (CoG) of the brain and the head are shown by the red and black points, respectively. At the ground FoR (A), the head deforms the brain by moving; at the skull FoR(B), the skull is fixed, and the brain is deformed by the inertial force by linear acceleration (Lin. Acc., $F_{LinAcc}$ in Eq.4, shown by yellow arrows), the inertial force by angular velocity (Ang. Vel., $F_{AngVel}$ in Eq.5, shown by green arrows) and the inertial force by angular acceleration (Ang. Acc., $F_{AngAcc}$ in Eq.6, shown by the blue arrows). (C) The theoretical framework of mechanical analysis about how brain strain is caused at the skull FoR. The directions of the force are assuming the head rotates and translates in the sagittal plane. The torque by angular acceleration ($T_{AngAcc}$) is non-zero and will drive the rigid-body rotation of brain (RRB, quantitatively described by equivalent angle of brain, Equ. Angle, $\phi$), which is linearly correlated to brain strain (quantitatively described by 95th percentile of maximal principal strain, 95% MPS).

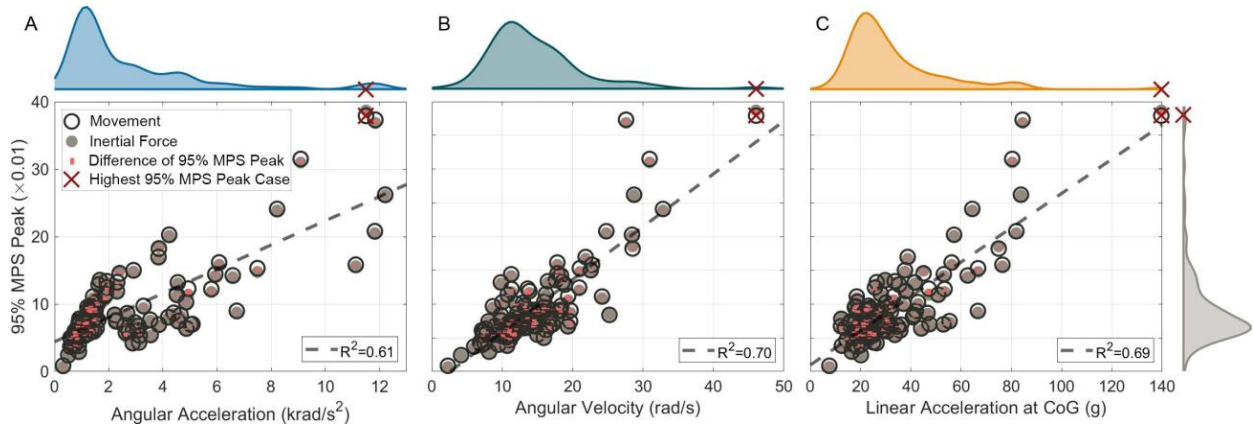

Figure 2. The kinematics and 95% MPS peaks in 118 on-field football data. 95% MPS peaks are calculated by the simulations with the loadings of head movement (black ring) and the inertial force (Eq.2, gray circle). The differences in the 95% MPS peaks given by two simulation methods are shown by the red bar. The case with the highest 95% MPS, which is used as an example in **Figs. 3-6**, is denoted by a red cross. The densities of distribution are plotted on the top and right of the figure. The blue, green, yellow and gray densities are for angular acceleration (A), angular velocity (B), linear acceleration at center of gravity (CoG) of the head (C), and 95% MPS peaks calculated by head movement, respectively. The 95% MPS peak by head movement is correlated to each kinematic parameter, and the $R^2$ is given at the right bottom.

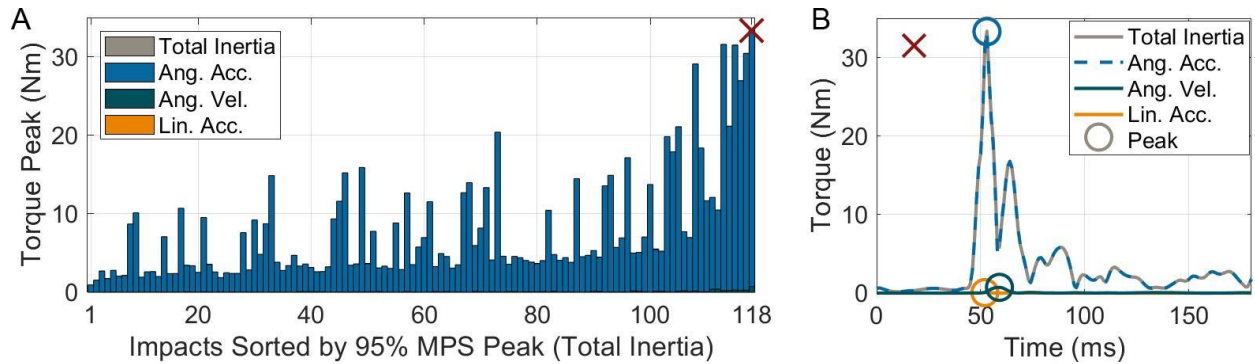

Figure 3. (A) Peak of inertial torques by all kinematics (Total inertia, Eq.8), angular acceleration (Ang. Acc., Eq.13), angular velocity (Ang. Vel., Eq.11) and linear acceleration (Lin. Acc., Eq.9). Each bar indicates one football head impact shown in **Fig.2**, and the impacts are sorted by the 95% MPS peak. (B) Traces of inertial torque by total inertia, angular acceleration, angular velocity, and linear acceleration in an example head impact. The head impact is denoted by red cross in (A). The peaks of the traces are indicated by the circles. The gray bars in (A) and curve in (B) for the total inertia is fully covered by the blue bars and curve for angular acceleration. The green bars for angular velocity and yellow bars for linear acceleration are near the bottom that can be barely seen.

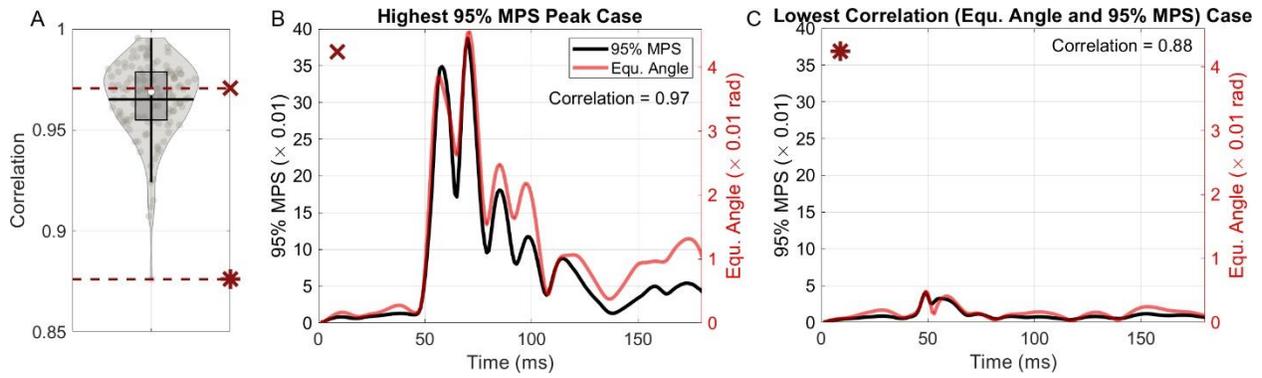

Figure 4. Comparisons between the equivalent angle of the brain $\phi$ and the 95% MPS. (A) The violin plots of Pearson correlation between $\phi$ and 95% MPS in 118 on-field football head impacts. (B, C) The trace of $\phi$ and 95% MPS in the head impacts with the highest 95% MPS peak (B) and the lowest correlation (C). The correlation is given at the right-top corner. These two cases are denoted by a red cross (Highest 95% MPS peak) and a red star (lowest correlation) in (A).

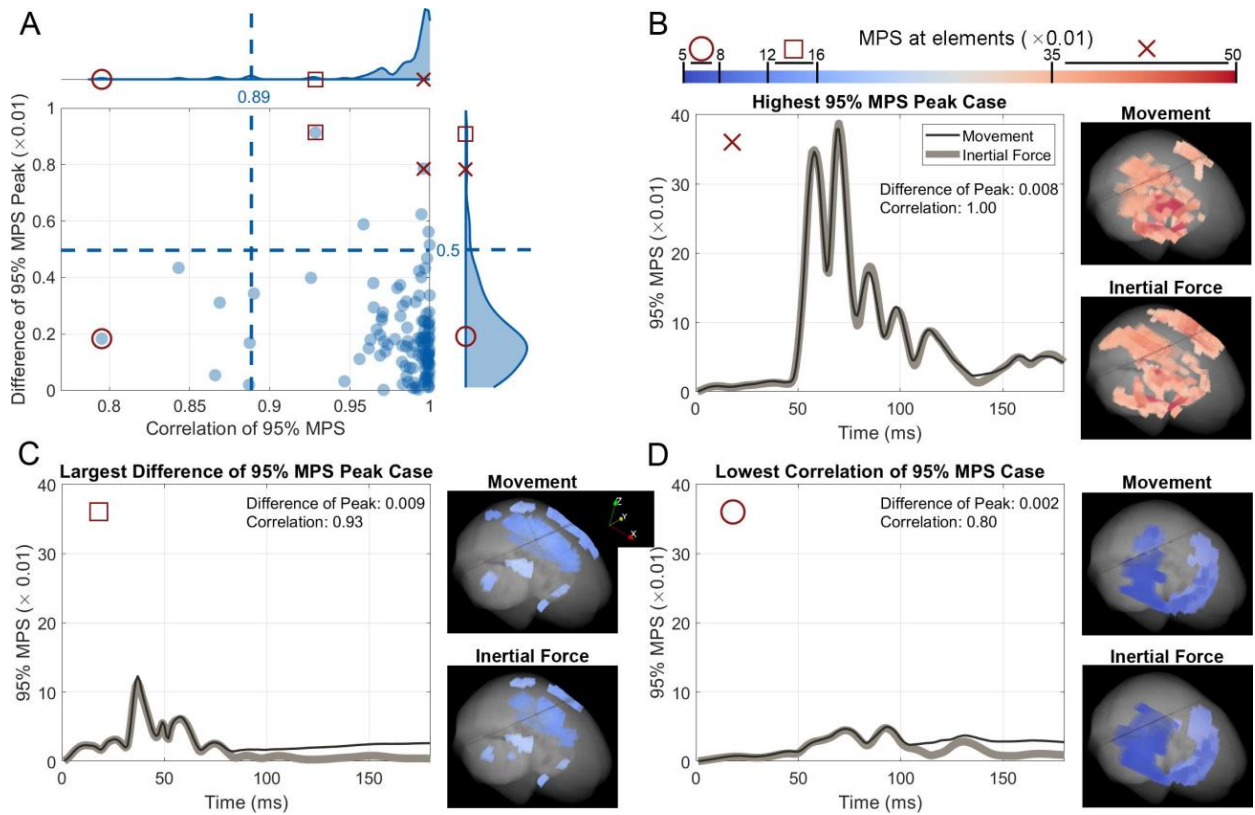

Figure 5. (A) The difference of the 95% MPS peak and the correlation of the 95% MPS traces calculated by head movement and inertial forces. Correlations in 95% head impacts are higher than 0.89, and the differences of the peaks in 95% head impacts are smaller than 0.005. The 95% MPS traces in the cases with the highest 95% MPS peak (B, red cross in A), the largest difference of peaks (C, red square in A), and the lowest correlation (D red circle in A) are plotted as examples, and the distributions of MPS when 95% MPS reached peak are given. The orientation of the brain is shown by the coordinates on the right of (C), X: back to front, Y: right to left, Z: bottom to top.

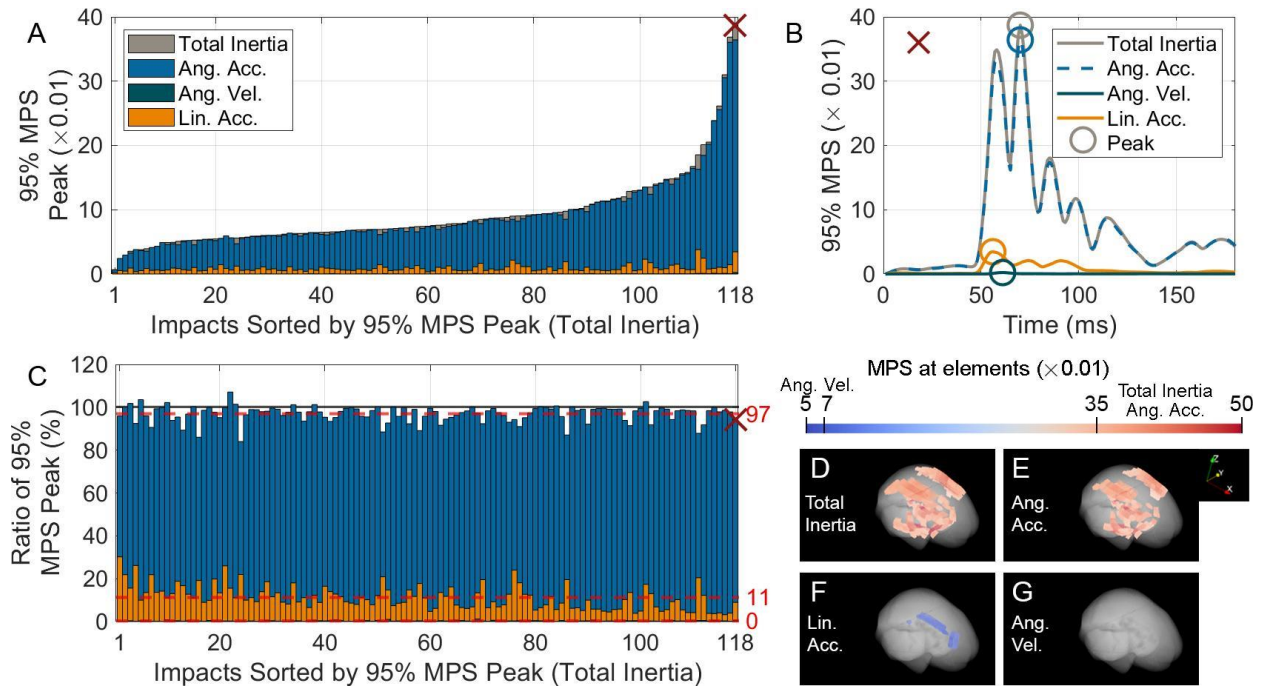

Figure 6. (A) The 95% MPS peak calculated by simulation with inertial force by all kinematics (Total Inertia, Eq.3), angular acceleration (Ang. Acc., Eq.6), angular velocity (Ang. Vel., Eq.5) and linear acceleration (Lin. Acc., Eq.4). (B) The 95% MPS traces calculated with different inertial forces in the case with highest 95% MPS peak (according total inertia). This case is denoted by red cross in (A, C). (C) The ratio of 95% MPS peak by individual kinematic parameter to 95% MPS peak by all kinematics. The mean ratio over 118 head impacts for angular acceleration, linear acceleration and angular velocity are 97%, 11% and 0%. (D-G): The MPS distribution when 95% MPS peak reach the maximum with total inertia (D), angular acceleration (E), linear acceleration (F) and angular velocity (G). The orientation of the brain is shown by the coordinates on the right of (E), X: back to front, Y: right to left, Z: bottom to top.

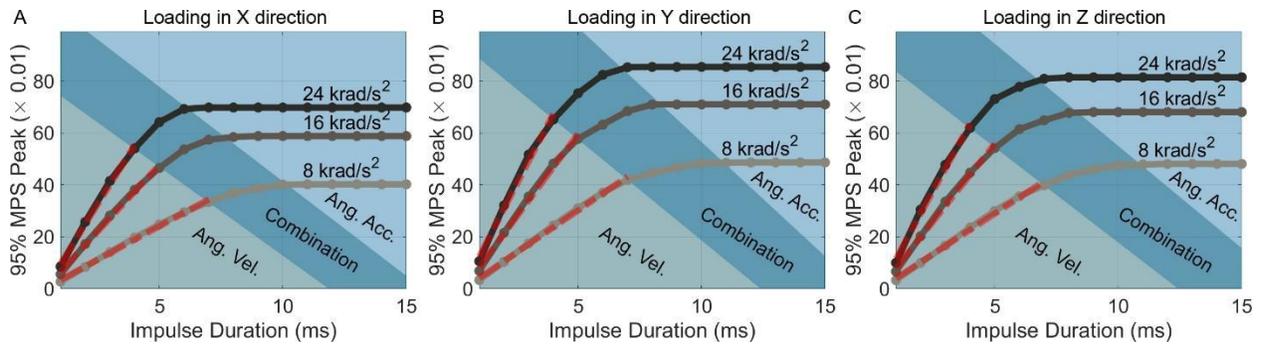

Figure 7. 95% MPS peak in simulations with constant angular accelerations of different impulse durations. Each curve corresponds to the same head angular acceleration (8, 16, 24 $\mathrm{krad/s^2}$) and each scatter corresponds to one simulation. (A) the constant angular acceleration in X direction (pointing front); (B) the constant angular acceleration Y direction (pointing right); (C) the constant angular acceleration Z direction (pointing bottom). In each plot, the left-bottom, right-top and the middle are the regions that 95% MPS peaks depend on head angular velocity peak, head angular acceleration peak and combination of them, respectively. In the angular velocity region, the curve is fitted to the linear line (red dot-dash), and the $R^2$ is all higher than 0.997.

Table.1 Abbreviations and symbols (**bold** means the variable is 3-dimensional vector)

| Abbreviation / Symbols | Meaning |
|---|---|
| mTBI | Mild traumatic brain injury |
| 95% MPS | 95th percentile maximal principal strain |
| RRB | Rigid-body rotation of the brain |
| RTB | Rigid-body translation of the brain |
| FE | Finite element |
| FoR | Frame of reference |
| CoG | Center of gravity |
| $\mathbf{r}$ | Position vector |
| $\mathbf{a}$ | Linear acceleration |
| $\boldsymbol{\omega}$ | Angular velocity |
| $\boldsymbol{\beta}$ | Angular Acceleration |
| $\mathbf{F}_{\text{LinAcc}}(\mathbf{r}), \mathbf{T}_{\text{LinAcc}}(\mathbf{r}), \mathbf{R}_{\text{LinAcc}}(\mathbf{r})$ | Inertial force, torque, resultant force at $\mathbf{r}$ by linear acceleration |
| $\mathbf{F}_{\text{AngVel}}(\mathbf{r}), \mathbf{T}_{\text{AngVel}}(\mathbf{r}), \mathbf{R}_{\text{AngVel}}(\mathbf{r})$ | Inertial force, torque, resultant force at $\mathbf{r}$ by angular velocity |
| $\mathbf{F}_{\text{AngAcc}}(\mathbf{r}), \mathbf{T}_{\text{AngAcc}}(\mathbf{r}), \mathbf{R}_{\text{AngAcc}}(\mathbf{r})$ | Inertial force, torque, resultant force at $\mathbf{r}$ by angular acceleration |
| $\mathbf{T}_{\text{skull}}, \mathbf{R}_{\text{skull}}$ | Torque, resultant force on brain by skull |
| $\mathbf{L}, \mathbf{P}$ | Angular momentum, momentum of brain in skull FoR |
| $\dot{\boldsymbol{\phi}}, \boldsymbol{\phi}$ | Equivalent angular velocity, angle of brain in in skull FoR |

Table.2 The range of the head impact kinematics.

| Sport/Source | Angular acceleration magnitude peak ($krad/s^2$) | Angular velocity magnitude peak ($rad/s$) | Linear acceleration at center of gravity of head magnitude peak (g) | mTBI included |
|---|---|---|---|---|
| Cadaver impact [36,37] | 0.8 – 22.4 | - | 12.1 – 163.7 | - |
| Voluntary head rotation [51] | 0.2 – 0.4 | - | - | No |
| Voluntary head rotation [50] | Quasi-statistic | - | - | No |
| Football [52, 32] | 0.2-14.4 | 4.1-48.0 | 5.1-162.4 | Yes |
| Football [53] | 1.1-10.9 | 6.6-64.5 | 18.8-138.2 | Yes |
| Football (Used as validation in letter) | 0.3-12.2 | 2.3-46.1 | 7.6-139.6 | No |
| Mix Martial Arts [54] | 0.5-10.7 | 4-35.0 | 7.1-175.3 | Yes |
| Mix Martial Arts [5] | 0.5-33.3 | 3.3-74.4 | 13.8-434.4 | Yes |
| Voluntary head rotation [28] | 0.1-0.5 | 2.7-24.7 | 1.6-13.0 | No |

# Supplementary Information

## S1. Inertial torques by angular acceleration and angular velocity on a spherical brain

In this section, we show how to calculate the integrations in Eqs.12 and 14 on a spherical brain, and how the Eqs.15 and 16 are obtained. Based on a Cartesian coordinate $(x, y, z)$ with origin at the center of the sphere, position vector $\boldsymbol{r}$, angular velocity $\boldsymbol{\omega}$ and angular acceleration $\boldsymbol{\beta}$ can be written as,

$$\boldsymbol{r} = (r_x, r_y, r_z) \qquad \text{(Eq.S1)}$$

$$\boldsymbol{\omega} = (\omega_x, \omega_y, \omega_z) \qquad \text{(Eq.S2)}$$

$$\boldsymbol{\beta} = (\beta_x, \beta_y, \beta_z) \qquad \text{(Eq.S3)}$$

For brevity, we give the inertial torques in $x$ directions as an example, and the inertial torque in $y$ and $z$ directions can be obtained by substituting the subscripts $(x, y, z)$ as $(y, z, x)$ and $(z, x, y)$. Assuming the homogeneous density $\rho_0$, replace Eqs. S1-S3 into Eqs. 13 and 14,

$$(\boldsymbol{T}_{\text{AngVel}})_x = \rho_0 \iiint (\omega_x r_z - \omega_z r_y)(\omega_x r_x + \omega_y r_y + \omega_z r_z) \cdot \mathrm{dv} \qquad \text{(Eq.S4)}$$

$$(\boldsymbol{T}_{\text{AngAcc}})_x = \rho_0 \iiint \left(\beta_x (r_x^2 + r_y^2 + r_z^2) - r_x(\beta_x r_x + \beta_y r_y + \beta_z r_z)\right) \cdot \mathrm{dv} \qquad \text{(Eq.S5)}$$

Then, we transform Eqs.S4, S5 to a spherical coordinate $(l, \theta, \xi)$ with the same origin as the Cartesian coordinates. Since $\boldsymbol{\omega}$ and $\boldsymbol{\beta}$ are unchanged in the integration, we only transform $\boldsymbol{r}$ as,

$$r_x = l \cos(\xi) \cos(\theta) \qquad \text{(Eq.S6)}$$

$$r_y = l \cos(\xi) \sin(\theta) \qquad \text{(Eq.S7)}$$

$$r_z = l \sin(\xi) \qquad \text{(Eq.S8)}$$

Replace Eqs.S6-S8 into Eqs.S4, S5,

$$(\boldsymbol{T}_{\text{AngVel}})_x = \rho_0 \iiint l^2 \big(\omega_y \cos(\xi) - \omega_z \sin(\xi) \sin(\theta)\big)\big(\omega_z \cos(\xi) + \omega_x \cos(\theta) \sin(\xi) + \omega_y \sin(\xi) \sin(\theta)\big) \mathrm{dv} \qquad \text{(Eq.S9)}$$

$$(T_{\text{AngAcc}})_x = \rho_0 \iiint -(\beta_x \cos^2(\theta)\sin^2(\xi) - \beta_x + \beta_z \cos(\xi)\cos(\theta)\sin(\xi)$$
$$+ \beta_y \cos(\theta)\sin^2(\xi)\sin(\theta))l^2 dv \quad \text{(Eq.S10)}$$

Then, integrate Eqs.S9, S10 in $l = [0, r_0]$, $\theta = [0, 2\pi]$, $\phi = [0, \pi]$,

$$\left(T_{\text{AngVel}}^{\text{Sphere}}\right)_x = 0 \quad \text{(Eq.S11)}$$

$$\left(T_{\text{AngAcc}}^{\text{Sphere}}\right)_x = \frac{8\pi}{15} r_0^4 \beta_x \quad \text{(Eq.S12)}$$

Similar deduction in Eqs.S4-S12 can be also performed in the $y$ and $z$ directions.

## S2. Inertia tensor

For one brain, momentum of inertia is different about different axis. About the axis ($\bar{n}$, a vector with unit length) passing the center of the gravity of the brain, the moment of inertia is,

$$I_C = \bar{n} \cdot I \cdot \bar{n} \quad \text{(Eq.S13)}$$

Where $I$ is the inertia tensor of the brain, and can be calculated by,

$$I = \begin{bmatrix} \iiint (r_y^2 + r_z^2)\rho(r)dv & \iiint -r_x r_y \rho(r)dv & \iiint -r_z r_x \rho(r)dv \\ \iiint -r_x r_y \rho(r)dv & \iiint (r_z^2 + r_x^2)\rho(r)dv & \iiint -r_y r_z \rho(r)dv \\ \iiint -r_z r_x \rho(r)dv & \iiint -r_y r_z \rho(r)dv & \iiint (r_x^2 + r_y^2)\rho(r)dv \end{bmatrix} \quad \text{(Eq.S14)}$$

Eq.21 shows that the direction of $L$, which is also the axis of $I_C$, is the same as $\omega$. Therefore, the $\bar{n}$ in Eq.S13 should be,

$$\bar{n} = \omega/|\omega| \quad \text{(Eq.S15)}$$

Where $|\omega|$ is the magnitude of $\omega$.